\documentstyle[12pt,aasms4,psfig]{article}

\slugcomment{to appear in the Astrophysical Journal}  
 
\begin{document}
 
\title{Centimeter searches for molecular line emission from
high-redshift galaxies} 
 
\author{
C. L. Carilli$^{1}$ \&
A.W. Blain$^{2}$
}          

\affil{$^{1}$National Radio Astronomy Observatory, P.O. Box O,
Socorro, NM 87801, USA, ccarilli@nrao.edu}      

\affil{$^{2}$Dept. of Astronomy, California Institute of Technology,
Pasadena, CA 91125, USA, awb@astro.caltech.edu}

\begin{abstract}

We consider the capabilities for detecting low order CO emission lines
from high-redshift ($z$) galaxies using the next generation of radio
telescopes operating at 22 and 43\,GHz. Low order CO emission studies
provide critical insight into the nature of high redshift galaxies,
including: (i) determining molecular gas masses, 
(ii) study of large scale structure through 
3-dimensional redshift surveys over cosmologically relevant
volumes, (iii) imaging gas kinematics on kpc-scales,  and, in conjunction
with observations of higher order transitions using future millimeter
telescopes, (iv) constraining the excitation conditions of the gas.
Particular attention is paid to the impact on such studies of the high
frequency limit for future centimeter telescopes.  We employ models
for the evolution of dusty star forming galaxies based on source counts
at (sub)millimeter (mm) wavelengths, and on the observed mm through
infrared (IR) backgrounds, to predict the expected detection rate of
low-order CO(2-1) and CO(1-0) line emitting galaxies for optimal
centimeter(cm)-wave surveys using future radio telescopes,
such as the Square Kilometer Array (SKA) and
the Expanded Very Large Array (EVLA). We then compare these results to
surveys that can be done with the next-generation mm-wave telescope,
the Atacama Large Millimeter Array (ALMA).  
Operating at 22\,GHz the
SKA will be competitive with the ALMA in terms of the detection rate
of lines from high-$z$ galaxies, and will be potentially superior by
an order of magnitude if extended to 43\,GHz. 
Perhaps more importantly, cm-wave telescopes
are sensitive to lower excitation gas in higher redshift galaxies, and
so provide a complementary view of conditions in high redshift
galaxies to mm-wave surveys. We have also included in our models 
emission from HCN. The number of HCN(1-0) detections 
will be about 5$\%$ of the CO detections in the (CO-optimized) 22 GHz
surveys, and about 1.5$\%$ for 43 GHz surveys.  
In order not to over-resolve the sources, brightness
temperature limitations require that a future large area
cm telescopes have much of its
collecting area on baselines shorter than 10 km.

\end{abstract}
 

\section {Introduction}

Observations of CO line emission from high redshift ($z$) galaxies
has become an important diagnostic tool in the study of galaxy
formation.  Such studies include observations of IR-luminous starburst
galaxies selected at IR and (sub)mm wavelengths (Brown \& van den Bout
1992; Solomon, Downes \& Radford 1992; Frayer et al.\ 1998; 
Ivison et al.\ 2001), and 
active galaxies (QSOs and radio galaxies) selected at optical
wavelengths (Omont et al.\ 1996a,b, 2001; Ohta et al.\ 1996; Barvainis et
al.\ 1998; Guilloteau et al.\ 1997, 1999; Papadopoulos et al.\ 2000,
2001; Carilli,  Menten \& Yun 1999; Carilli et al.\ 2002).  Due to the
sensitivity limitations of current mm-wave telescopes, such studies
have been limited to extreme high mass, high luminosity galaxies,
corresponding to galaxies with far-IR (FIR)
luminosities: $L_{\rm FIR} \ge 10^{12}$ L$_\odot$, and molecular gas
masses: $\rm M(H_2) \ge 10^{10}$ M$_\odot$. While such systems are
rare in the nearby universe, 
$10^{-6}$ Mpc$^{-3}$, or 10$^{4}$ times lower than $L_*$ galaxies,
(sub)mm surveys suggest that the comoving number density of
such objects increases by up to three orders of magnitude by $z
\sim 2$ to 3 (Smail, Ivison \& Blain 1997; Blain et al.\ 1999; Barger
et al. 1998). Another serious limitation to current instruments
is the narrow bandwidth of the spectrometers. Current spectrometers
have maximum bandwidths of 0.5 to 1 GHz, corresponding to 1500 to 
3000 km s$^{-1}$ at 100 GHz.  The limited sensitivity and bandwidth of
existing instruments effectively preclude large-volume searches for
high-$z$ galaxies via their molecular line emission.

The next generation of mm- and cm-wave telescopes
will have sensitivities and bandwidths at least an order of
magnitude better than existing instruments. With such
capabilities it becomes feasible to consider searches for high-$z$
galaxies at mm and cm wavelengths.
This problem has already been addressed in detail for
(sub)mm-wave observations using the   Atacama Large Millimeter Array
(ALMA), the next-generation mm-wave telescope
(Blain et al.\ 2000 -- Paper I; Blain 1996, 2001; van der Werf \&
Israel 1996;  Silk \& Spaans 1997; Combes, Maoli, \& Omont
1999;  Gnedin, Silk \& Spaans 2001). 

This paper dramatically expands the discussion in Paper I of the 
capabilities of future telescopes to study the
evolution of dusty star forming galaxies via their low-excitation 
molecular line emission. In Paper I, formalism was developed to 
investigate optimal molecular line surveys which maximize the 
detection rate of galaxies for a given amount of observing time,
and the analysis presented herein relies on the same formalism.
The importance of such surveys is delineated in detail
in Paper I. To summarize, measuring molecular lines, and
especially the  low-excitation lines that trace the bulk of gas in the
interstellar  medium (ISM), provides a direct probe of the massive gas 
reservoirs required to fuel star formation in nascent
galaxies. In addition, the velocity width of the lines provides an 
estimate of the dynamical mass of the systems.   
Observations of the infrared background, and
source counts at (sub)mm wavelengths, suggest that a
substantial fraction of the star formation that occurs in the universe
is invisible at optical wavelengths due to dust obscuration (Blain et al.\
1999). Future
molecular line surveys may be the only way of obtaining a complete
census of the redshift distribution of this cosmic star formation.

Paper I emphasized the capabilities of future mm-wave
telescopes, such as the ALMA, for discovering high redshift
galaxies via their CO emission.  The focus of the
analysis presented herein is on 
future cm-wave telescopes, such as the Square Kilometer Array (SKA)
and  the Expanded Very Large Array (EVLA), for observing 
low-excitation CO molecular line emission from
high-$z$ galaxies. The important difference between mm- and cm-wave
observations is that mm-wave telescopes are limited to studying higher
order CO transitions from high-$z$ galaxies -- CO(3-2) or higher at
100\,GHz for $z>2$ -- while cm-wave telescopes probe the lower order
transitions, CO(2-1) and CO(1-0). Clearly such complementarity is
critical for determining excitation conditions in high-$z$ molecular
gas clouds. In particular, mm-wave surveys of high order transitions
will be biased toward
high-density, high-temperature gas, with $n({\rm H}_2) >
10^3$\,cm$^{-3}$ and $T>30$\,K (Papadopoulos et al.\ 2001;
Papadopoulos \& Ivison 2001).  

An important advantage of cm-wave surveys is the 
large fractional bandwidth and primary beam area.
At high redshifts cm-wave spectroscopy will
be able to map out the 3-dimensional spatial structure of line-emitting 
galaxies, and thus to trace directly the build up of large-scale structure 
in the Universe. Both the thickness in redshift, and the projected area on   
the sky of the surveyed volume, will be 
much greater in the cm waveband than at mm wavelengths. It is also likely 
that low-excitation lines will map out a larger fraction of the volume of 
the ISM in these galaxies, and thus provide the opportunity to study in 
detail the spatially resolved kinematic structure of  
most of the gas in the ISM in high-redshift galaxies, which resides in 
the cold phase. These observations would provide information about both  
the evolution of the masses of galaxy disks and the disruption of disks by 
galaxy interactions at large redshifts. This will allow valuable new tests 
of understanding of the assembly and formation of galaxies. 

We use the latest galaxy evolution models from Paper I, modified to
take account of additional data and  
revised cosmology. While these models 
represent perhaps the best current estimates
of the evolution of dusty star forming galaxies, the 
observational constraints are such that substantial revisions,
especially  of the source redshift distribution, 
are certainly possible. Likewise, the design specifications
of the next-generation cm-wave telescopes are
still being considered. Hence, this paper is not meant
as a definitive and final prediction of the results of 
future molecular line surveys
with cm-wave telescopes, but merely to advocate the 
possibilities. A particular concrete question we hope to address is:
what choices in the design of
such telescopes (eg. high frequency limit, maximum baseline) will
facilitate the study of molecular line emission from  high-$z$
galaxies? We assume H$_0$ = 65 km s$^{-1}$ Mpc$^{-1}$,
$\Omega_M = 0.3$, and $\Omega_\Lambda = 0.7$ 

\section{Dusty galaxy evolution models}

The models we employ for predicting CO source counts are updated 
from Paper I. These models employ an analytic description of 
pure luminosity evolution of the low-$z$ {\it IRAS} 60-$\mu$m luminosity 
function (Saunders et al.\ 1990). The evolution function has the form, 
\begin{equation}
g(z) = (1+z)^{3/2} {\rm sech}^2 [ b {\rm ln}(1+z) - c ] \, {\rm cosh}^2 c.
\end{equation}
At very low and very high redshifts the function can be 
approximated by $g(z) \propto (1+z)^{\gamma}$, with $\gamma \simeq 
{3/2} + 2b\,{\rm tanh}\,c$ and $\gamma = {3/2} - 2b$ respectively.  
With the current cosmology, and taking into account all available far-IR 
and (sub)-mm background, count and redshift distribution data, the values 
$b = 2.2 \pm 0.1$ 
and $c = 1.84 \pm 0.15$ are required, if a typical dust temperature of 37\,K 
is assumed. The fitting procedures and details of the information used 
are explained in Blain et al.\ (1999). A plot of this function is shown 
in Fig.\,1 of Blain (2002).  
The evolution peaks at $z \simeq 1.7$, at which the bolometric luminosity 
density of infrared-luminous galaxies is 
about 40 times greater than at $z=0$. 

The strength of the emission from CO lines is calculated as described in 
Paper I. Excitation conditions, and thus line ratios, are derived from
a  standard large velocity 
gradient model (Frayer \& Brown 1997) with a 
kinetic temperature of about 50\,K and a density of 10$^4$\,cm$^{-3}$. This 
model provides a reasonable description of the CO emission from two 
well-studied FIR-luminous galaxies at $z \sim 2.5$, and their properties 
are used to normalize the CO emission line strengths to the evolving 
bolometric luminosity function of galaxies described above. 

\section{Telescope parameters}

Table\,1 lists the assumed parameters for telescope
capabilities for the
SKA,\footnote{http://www.skatelescope.org} 
EVLA,\footnote{http://www.aoc.nrao.edu/doc/vla/EVLA} and 
ALMA.\footnote{http://www.alma.nrao.edu/info}   
Columns 2 and 3 give the antenna diameter
and the instantaneous frequency range covered, respectively.  Column 4
gives the effective aperture (total collecting area $\times$
aperture efficiency) divided by the system temperature,
and Column 5 gives the Field of View (FoV) of the
primary elements at the given frequencies. 
Column 6 gives the rms sensitivity in 1 hour for
a FWHM spectral channel of 300\,km\,s$^{-1}$. 

The design for the SKA, including the primary stations for the array,
the correlator, and the frequency range covered, is still under
investigation.  For the analysis below we adopt the United States SKA
consortium design concept of a square kilometer of collecting area
comprised of small (7m) diameter antennas, arranged in 1000 stations
of 26 antennas per station (Cordes, Preston, \& Tarter 2001).  
We assume the elements in each station are
closely packed ($\sim$ 1m separation), implying an effective station
diameter of about 40m.  Beam-forming correlators generate `effective
beams' at each station corresponding roughly to the diffraction
limited beam of 
a 40m diameter antenna. Signals from these effective beams are then
cross correlated with the corresponding beams of other stations.  The
current specifications for the SKA require imaging of the full primary
beam of the 7m elements, thereby requiring roughly
$({{40}\over7})^2$ = 33 beams per station.  The number of cross
correlations (per polarization and lag channel) required for this type
of a system is then the sum of the number of cross correlations
between stations $= {{1000^2}\over2} \times 33 = 1.6\times10^7 $, 
plus the number
of cross correlations per station (ie. the beam-forming correlator) 
$ = {{25^2}\over2} \times 1000 \times 33 = 1.0 \times 10^7$, for a
total of $2.6 \times 10^7$ cross correlations.  This value is an order
of magnitude smaller than the $3.4\times10^8$ cross correlations
required in the case of full cross correlation of the 26000 7m
elements.  The 
trade-offs between full cross correlation and beam forming instruments
are in imaging fidelity, which is not a critical issue for the CO
searches discussed herein, and in calibration flexibility, ie. the
difficulty and complexity is shifted from the construction of the 
large correlator required for full cross
correlation to ensuring the stability of the many beam-forming correlators.  
The analysis discussed below is not critically
dependent on the specific design of the array stations, except with
respect to the imaging FoV.  For a fixed collecting area a large FoV
is clearly preferable for surveys, as the source detection rate
increases linearly with FoV.

An important point is the high-frequency limit of the SKA.  The
current straw-man design has a maximum frequency of 22\,GHz. 
However, Weinreb \& Bagri 
(2001) have shown that going to higher frequencies may be feasible,
both in terms of the  antennas and the receivers. In this paper
we consider CO line searches in a 43-GHz band in addition to the
standard 22-GHz band.  One of the main motivations of this paper is to
gauge the scientific advantages of going to higher frequency with 
future large area cm telescopes.
Such information is required for proper understanding of the
trade-offs between scientific capability and cost.
We assume a 4-GHz instantaneous
total bandwidth, and a 22-GHz aperture efficiency that is a factor two
less than that at 5\,GHz, ie. a `half-SKA'.

The EVLA is included in the analysis since it represents the
nearer-term capabilities for cm-wave 
high-$z$ CO surveys.  The EVLA represents a major step forward
in the study of low-order high-$z$ molecular lines
in a number of ways. First, through improved receivers and 
antenna structures the EVLA will be about twice as sensitive 
as the current VLA at 43\,GHz for 
spectral line observations. Second, the current VLA
correlator is limited to only a 50-MHz bandwidth
with 7 spectral channels, which corresponds 
to only 350\,km\,s$^{-1}$ at 43\,GHz. Such a narrow band both 
precludes searches for high-$z$ CO emission
and provides very limited spectral information (Carilli et al.\ 1999).
The EVLA will have an 8-GHz bandwidth in two polarizations
with 16000 spectral channels. 
Third, the current VLA high-frequency bands are limited
to 20.5--25~GHz and 40--49\,GHz.
The EVLA will have continuous frequency coverage from 1 to 50\,GHz. 

The ALMA is included as a standard of comparison for future (sub)mm
telescopes (see Paper I for details).  For both the EVLA
and the ALMA we consider the optimum bands for CO searches (40
and 230\,GHz,  respectively).  
Note that the sensitivity of the EVLA at 43 GHz is
comparable to that expected for ALMA at 43 GHz -- the larger collecting
area of the EVLA is off-set by the higher aperture efficiency and
better observing site for ALMA. The inclusion of a 43 GHz system 
for ALMA is currently under debate. If such a system is included,
then the sensitivity for the EVLA in Table 1 is essentially
that expected for the ALMA at this frequency, while
the ALMA will have  a factor four larger FoV. 

Arrays of heterodyne receivers on large single dish
telescopes offer a potentially competitive method for 
performing wide field  molecular line surveys.
In order to sample the same area of the sky with a single dish
relative to an interferometer
the number of independent single dish 
beams that need to be sampled is given by the ratio of the FoV of the
interferometer to that of the single dish. For example,
the 100-m Green Bank Telescope\footnote{http://www.gb.nrao.edu/GBT/} 
and the EVLA have similar collecting areas, so a 16 element receiver
array on the GBT will survey a similar area of the sky
as the EVLA with comparable sensitivity.  
A similar size receiver array on the 50-m Large Millimeter Telescope
(LMT)\footnote{http://lmtsun.phast.umass.edu/} will sample
the same FoV as the ALMA, with about 25$\%$ of the
sensitivity.  Of course, a wide-band autocorrelation spectrometer will
be required for each element of the receiver array. The telescope
optics in both cases  (the GBT and the LMT) 
will support such receiver arrays, since 
both   telescopes have been designed to
support large format ($30\times30$) bolometer cameras.

\section{Analysis}

\subsection{Optimal Surveys}

As a metric of the capabilities of the different telescopes
we use the detection rate for an `optimal survey'.
The optimal survey balances area surveyed versus
sensitivity to sample the source counts at the `knee' of the
distribution, corresponding to the regime where
$N(>S) \propto S^{-2}$, where $S$ is the flux in
the line (see Paper I for details). This criterion maximizes the
number of sources detected in a given observing time. 

In Fig.\,1 we show the predicted 22- and 43-GHz line source
counts for a 4-GHz bandwidth using the formalism for
infrared-selected  galaxy evolution developed in Paper I.
The line counts increase approximately linearly with bandwidth. 
Table 2 lists the frequency ranges and optimal depths 
for the different telescopes, as well as the number of pointings per 
hour for the optimal depth and the detection rate of high-$z$
CO-emitting galaxies. For the baseline specification of a 22 GHz
high frequency upper limit,  the SKA galaxy detection rate
will be a factor two larger than that of the ALMA. If the high
frequency limit is increased to 43 GHz, then the SKA becomes the
pre-eminent instrument for discovering high redshift galaxies via
their CO emission, detecting 100's of galaxies per hour. 
The EVLA is clearly much slower than the SKA in terms of
discovering high redshift galaxies 
due to its lower sensitivity and smaller FoV. 

It is important to keep in mind that the line surveys at 
different frequencies are sampling different
CO transitions at different redshifts. 
In Fig.\,2 we delineate the detection rates in terms
of the different transitions and redshifts. 
Surveys with the SKA at 22\,GHz to the optimal depth of
$10^{-23}$\,W\,m$^{-2}$  
detect almost exclusively the CO(1-0) line from galaxies with 
$L_{\rm{FIR}} \ge 1.6 \times 10^{12}$\,L$_\odot$ 
at $z \simeq 4.2$. At 43\,GHz at the optimal depth of
$10^{-22}$\,W\,m$^{-2}$,  about 92\% of the detections are CO(1-0)
lines from galaxies with  
$L_{\rm{FIR}} \ge 2\times 10^{12}$\,W\,m$^{-2}$ at
$z \simeq 1.7$, with about 7.5$\%$ being CO(2-1) lines from 
galaxies with  $L_{\rm{FIR}} \ge 2.4\times 10^{12}$ 
at $z \simeq 4.3$, and
perhaps 0.5\% of the detections being CO(3-2) emission
from galaxies with similar
FIR luminosity but at  $z \simeq 7.0$, if such galaxies exist. 

Surveys with the ALMA at 230\,GHz to the optimal depth of 
$4 \times 10^{-21}$\,W\,m$^{-2}$ are dominated by galaxies with 
$L_{\rm{FIR}} \ge 4\times 10^{11}$\,L$_\odot$, 
at $z \sim 1$ to 2,
emitting CO(4-3) to CO(6-5), with 
a minor contribution from galaxies with 
$L_{\rm{FIR}} \ge 5\times 10^{10}$\,L$_\odot$ at $z \sim 0.5$ 
emitting CO(3-2) and from more-luminous galaxies with 
$L_{\rm{FIR}} \ge 2.4 \times 10^{12}$\,L$_\odot$ emitting 
CO(7-6) at $z \sim 2.5$.  The
ALMA can also potentially detect a 
comparable number of redshifted 
fine-structure lines from higher redshifts in this band
(Paper I). 

\subsection{Other issues: Brightness temperature, clustering, and HCN
contamination} 

We consider briefly a few issues relating to the cm-wave study of
CO emission from high redshift galaxies, including: (i) maximum
baselines for the array given the expected brightness temperatures,
(ii) the capabilities of studying large scale structure through
optimal surveys, and (iii) contamination of such surveys by 
emission from HCN -- the  next strongest thermal molecular emission
species in the relevant frequency range. 

An important issue when considering detecting thermal emission
from high redshift galaxies with future cm telescopes is the
maximum baselines dictated by the intrinsic brightness
temperature of the emission.  The expected (rest-frame) brightness
temperature for the CO emission from high-$z$ galaxies is likely to 
be $\le 40$ K. This corresponds to an observed brightness temperature
of $\le 8$ K for a source at $z = 4$. 
The peak flux density of the line for the optimal 43-GHz survey 
corresponds to about 0.2\,mJy for the SKA. In order not to spatially
over-resolve the sources, much of the
collecting area of the array must then be on baselines $\le 10$\,km,
corresponding to $\ge 0.14''$ resolution at 43 GHz. 

The second issue is large scale structure. 
The large number of galaxies  discovered over a
relatively narrow redshift range for sensitive 
molecular line surveys at cm wavelengths has the added
benefit of facilitating 
three-dimensional studies of large-scale structure 
at moderate and high redshifts.
Operating at 20--24\,GHz, the optimal SKA survey with a
station diameter of 7\,m detects CO(1-0)
emission from about 360 galaxies in 24\,hours. These
galaxies are in the redshift range $z=3.79$ to 4.75 over
an area of 3.5\,deg$^2$, corresponding to a comoving volume
of 0.044\,Gpc$^3$. At the higher frequency of 40-44\,GHz,
the CO(1-0) emission from about 4200 galaxies
would be detected in 24\,hours from $z=1.61$ to 1.88 over an
area of 23\,deg$^2$ and a comoving volume of 0.085\,Gpc$^3$.

A volume of about 0.1\,Gpc$^3$ corresponds to a representative
volume of the universe, and so these surveys will provide excellent 
probes of large-scale structure. The geometries of both of
these daily survey volumes are only about five times deeper than they
are wide. This should ensure that three-dimensional large-scale structure 
is sampled reliably. The space density of CO-emitting galaxies
is expected to be about 100 times lower than for optically-selected, 
spectroscopically-confirmed Lyman-break galaxies 
(Steidel et al. 1999), and about 10
times less than 
submm-selected galaxies (Blain et al. 1999). However, the geometry of
the three-dimensional redshift surveys should make low-excitation CO 
surveys 
an ideal tracer of large-scale structure. In particular, as
the CO line surveys sample higher luminosity
galaxies, their clustering strength relative
to the dark matter at high redshift might 
be  stronger than for lower
luminosity  galaxies in standard  hierarchical galaxy formation
models (Davis et al. 1985). Whether this is true will provide important 
new information about bias and the galaxy formation process, especially  
as mass measurements will be available for comparison from CO line  
widths and excitation ratios.  

Finally there is the issue of HCN emission.
Solomon et al. (1997) have shown a non-linear relationship between
$L'$(CO(1-0)) (in K km s$^{-1}$ pc$^2$) 
and $L_{\rm FIR}$  for galaxies in the sense that high $L_{\rm FIR}$ 
galaxies have lower values of $L'$(CO(1-0)) than would be expected
based on a linear relationship.
Considering $L_{\rm FIR}$  to be a measure of
star formation rate, and $L'$(CO(1-0)) to be a measure of molecular
gas mass, this would imply an increase in star formation efficiency
($\equiv \rm {{Star~Formation~Rate}\over{Gas~Mass}}$) with increasing
luminosity, in particular for galaxies with $L_{\rm FIR}$  $\ge 10^{11}$
L$_\odot$ (Solomon et al. 1997).  For dense nuclear starbursts a
number of groups (Solomon et al. 1997; Mao et al. 2000;
Weiss et al. 2001) have shown that the densities are such that the
entire interstellar medium in the starburst regions may be molecular,
and that the CO(1-0) emission may be dominated by this molecular
inter-cloud medium, as opposed to being from the denser star forming
clouds themselves.  In this situation it appears that higher density
molecular tracers, such as HCN, which has a critical density of order
$10^5$ cm$^{-3}$, may be a better probe of the star forming clouds
themselves.  Gao \& Solomon (2001) show that the relationship between
$L'$(HCN(1-0)) and $L_{\rm FIR}$  remains linear, with $L_{\rm FIR}  = 863
\times L'(HCN(1-0))$ over the range: $L_{\rm FIR} = 10^{10}$
L$_\odot$ to $10^{12}$ L$_\odot$ (see Solomon 2001).
Given this linear relationship, and 
the non-linear relationship between $L_{\rm FIR}$  and $L'$(CO(1-0)),
the ratio: ${{L'({\rm HCN(1-0)})}\over{L'({\rm CO(1-0)})}}$ varies
from about 0.025 for galaxies with 
$L_{\rm FIR}$  = 10$^{10}$ L$_\odot$ to
0.10 for  $L_{\rm FIR}$  = 10$^{12}$ L$_\odot$.

In order to estimate the contamination of optimal cm-wave CO
surveys by HCN emission, we have included the HCN(1-0) (88.6 GHz rest
frequency)  emission in the galaxy
formation models discussed in section 2. At the most efficient depths
for the CO searches,  we find that the number of  HCN(1-0) detections
would be about  5$\%$ of the total number of CO(1-0) lines at 22 GHz and about
1.5$\%$ of the total number of lines at 43 GHz. 
This emission would correspond to galaxies with $L_{\rm FIR}$  $\sim
10^{13}$ L$_\odot$ at $z \sim 3$ for the 22 GHz survey, 
and  similar luminosity galaxies at $z \sim 1$ for the 
43 GHz survey. Hence, HCN emission should not present a major
confusion problem for such optimal cm-wave CO line searches. 
On the other hand, we also find that the fractional contamination by HCN
is likely to increase with the depth of the survey, eg. a 22-GHz survey 
to a depth of 
$10^{-24}$ W m$^{-2}$ will contain about 20$\%$ HCN(1-0) lines. 

\section{Discussion}

From Table\,1 and Fig.\,2, it can been seen that future large area
cm-wave telescopes  operating at 22 GHz  will be
competitive with future mm-wave telescopes 
in terms of discovering high-$z$ star-forming
galaxies through their molecular line emission. Increasing the
high frequency limit to 43-GHz allows for optimal
surveys which are 20 times faster
than the ALMA in terms of discovering high-$z$ galaxies. 

Detection rates are a simplistic metric for the ALMA and the SKA, and
it is important to emphasize their complementarity. ALMA surveys will
be dominated by higher-order CO lines from intermediate redshift ($z
\sim 1 - 2$),   intermediate luminosity (L$_{\rm FIR} \sim ~ \rm
few\times10^{11}$ L$_\odot$) objects. Surveys with the SKA at 22 GHz
will be dominated by low-order
transitions from higher luminosity (L$_{\rm FIR} \sim 10^{12}$ L$_\odot$)
galaxies at higher  redshifts ($z \sim 4$).  

For the EVLA at 43 GHz
the combined small FoV and low sensitivity 
make it a much slower survey instrument that either the ALMA 
and SKA. However, the EVLA has adequate sensitivity 
to resolve and study at sub-arcsec angular resolution 
the low-order CO emission from individual
FIR-luminous, high-$z$
objects selected from wide-field surveys at other wavelengths, such
as surveys using (sub)mm bolometer arrays on large single-dish
telescopes. And perhaps most importantly,
the large fractional bandwidth of the EVLA
will allow for redshift determinations via molecular line 
searches using pointed observations of individual objects.
The EVLA will thus provide the first look into the nature
of low-order CO emission from high-$z$ galaxies. The
importance of such capabilities has already been 
demonstrated with the current VLA in a few extreme
cases, although the limited bandwidth
effectively precludes  proper spectroscopy 
(Papadopoulos et al.\ 2001; Carilli et al.\ 1999, 2002).

The CO excitation conditions assumed in the models of section 2 could
lead to pessimistic predictions for the low-order transitions
because the velocity integrated line flux density increases roughly
with the square of the 
frequency (ie. constant brightness temperature), at least to CO(4-3)
(Paper I). While this appears to be roughly appropriate for 
infra-red selected galaxy samples, our models do not include
the possibility of a population of molecular gas-rich, high redshift
galaxies with lower excitation conditions. For example, the
CO excitation conditions for the Milky Way disk inside the solar
radius (excluding the Galactic center)
imply roughly equal velocity integrated flux density for CO(1-0) and
CO(4-3) (Fixsen, Bennett, \& Mather 1999). If such a population of 
galaxies exists, then the predictions from our models
can be considered lower limits to the cm-wave source counts for
molecular line surveys. Recent observations with the VLA provide
evidence that such a population may indeed exist (Papadopoulos et al.\
2001). 

Limitations to the intrinsic brightness temperature of the thermal
line emission from high redshift galaxies require that much of the
collecting area of future large area radio telescopes be concentrated
on baselines $\le 10$ km.  On the other hand, having baselines out to
10 km provides the important capability of imaging the emission on
scales relevant to galaxies ($0.14'' \simeq 1$ kpc), and for resolving
the multiple images of the order of 1\% of line-emitting galaxies
expected to be gravitationally lensed by foreground galaxies.

Lastly, we have found that HCN(1-0) emission will not
be a major source of confusion to optimal cm-wave CO line searches,
comprising about 5$\%$ of the total number of detected galaxies
at 22 GHz and 1.5$\%$ at 43 GHz. Of course, once identified as such, 
HCN is interesting in it's own regard as a better tracer than CO of
the star forming clouds in active star forming galaxies (Solomon
2001). 

Like studies of star formation in our own galaxy, 
it has become clear that a complete census of the star-formation
history of the universe requires an understanding of the contribution
from galaxies that are obscured by dust at optical wavelengths. 
The next generation mm- and cm-wave telescopes will provide unique,
and complementary,  capabilities for studying the  thermal and
non-thermal line and continuum emission from such systems 
at sub-arcsecond spatial resolution. 

\vskip 0.2truein 

The National Radio Astronomy Observatory (NRAO) is operated by
Associated  Universities, Inc. under a cooperative agreement with the
National Science Foundation. We thank Dave Frayer for further
discussions  about the line models implemented in Paper I.

\newpage
 
\begin{table}
\caption{Telescope Parameters} 
\vskip 0.2in
\begin{tabular}{cccccc}
\hline
\hline
Telescope & Antenna diameter & Frequency & 
$A_{\rm{eff}}/T_{\rm{sys}}$ & FoV & 1-hr rms$^a$ \\ 
~ & (m) & (GHz) & (m$^2$\,K$^{-1}$) & (arcmin$^2$) & ($\mu$Jy) \\
\hline
SKA & 7 & 20 - 24 & 5600 & 40 & 2.5 \\
SKA & 7 & 40 - 44 & 4000 & 10 & 2.5 \\
EVLA & 25 & 38 - 46 & 75 & 0.8 & 100 \\
ALMA & 12 & 222 - 238 & 100 & 0.15 & 45 \\
\hline
\end{tabular}

~$^a$rms noise in 1 hour for a 300 km s$^{-1}$ channel.

\end{table}

\begin{table}
\caption{Detection rates for optimal surveys.} 
\vskip 0.2in
\begin{tabular}{cccccc}
\hline
\hline
Telescope & Frequency & Optimal depth & Pointings$^a$  & Rate$^b$ \\ 
~ & (GHz) & (W\,m$^{-2}$) & (hour$^{-1}$) & (hour$^{-1}$) \\
\hline
SKA &  20 - 24 & $10^{-23}$ & 13 & 15 \\
SKA &  40 - 44 & $10^{-22}$ & 346 & 176 \\
EVLA & 38 - 46 & $10^{-22}$ & 0.22 & 0.02 \\
ALMA & 222 - 238 & $4\times10^{-21}$ & 60 & 7.5 \\
\hline
\end{tabular}

~$^a$Number of pointings per hour for the optimal survey. 

~$^b$Number of 5$\sigma$ line
sources detected per hour observing time.

\end{table}

\clearpage
\newpage

\centerline{Figure Captions}

\noindent {\bf Figure 1}: Cumulative source counts of 300-km-s$^{-1}$
wide CO lines at 22\,GHz (solid line) 
and 43\,GHz (dash line), developed from the results of 
Blain et al.\ (2000, Paper I). A 4-GHz-wide band is assumed. 

\noindent {\bf Figure 2}: The detection rate of high-$z$ CO emission
lines for `optimal surveys' using
the ALMA at 230\,GHz and the SKA at 22 and 43\,GHz.
The histograms delineate the contribution from different
CO transitions at different redshifts, 
as listed in each segment of the histogram. Note that the quoted redshifts
are the mean values for the band covered.

\clearpage
\newpage

\begin{figure}
\psfig{figure=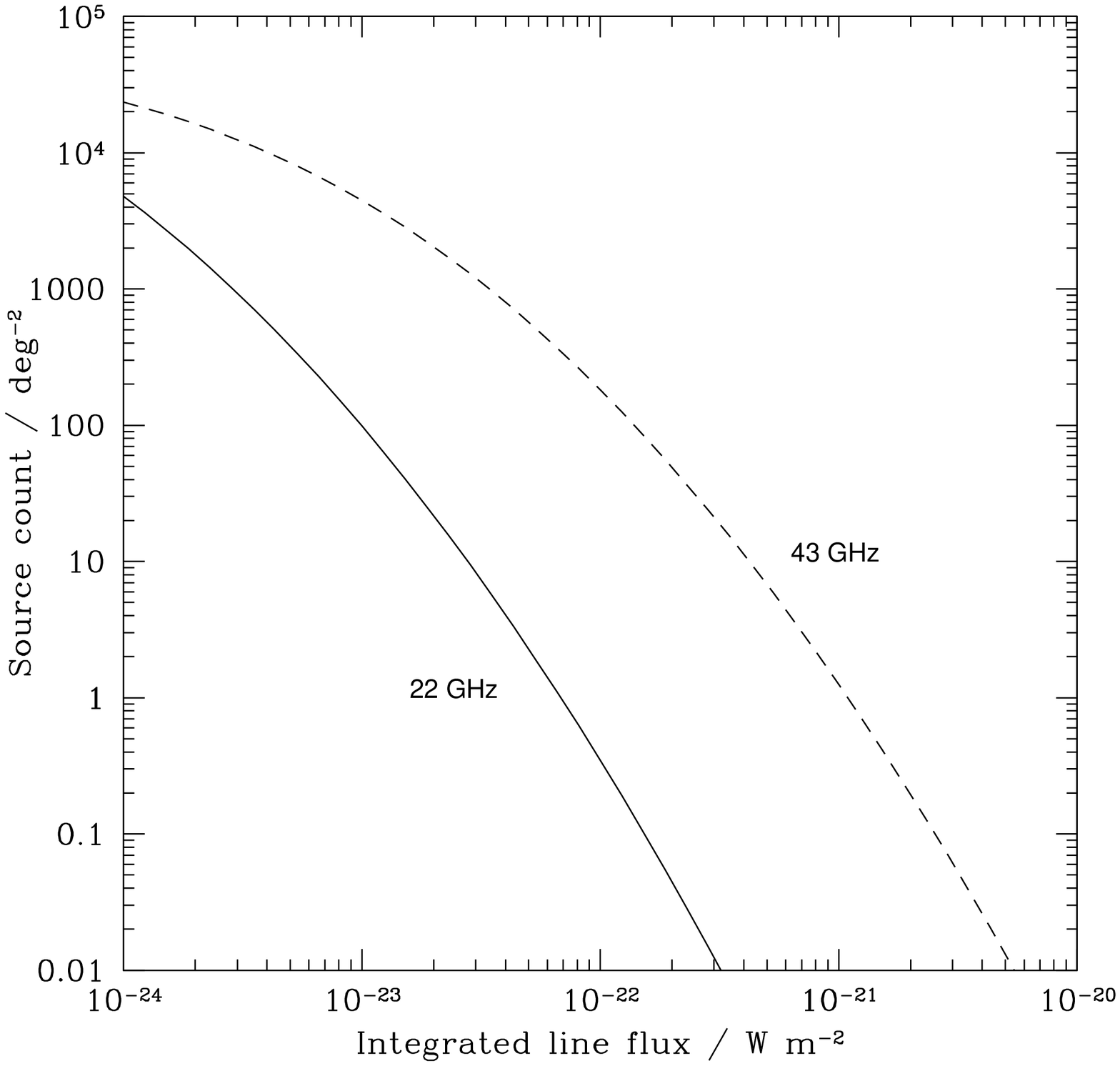,width=6in}
\caption{}
\end{figure}

\clearpage
\newpage

\begin{figure}
\psfig{figure=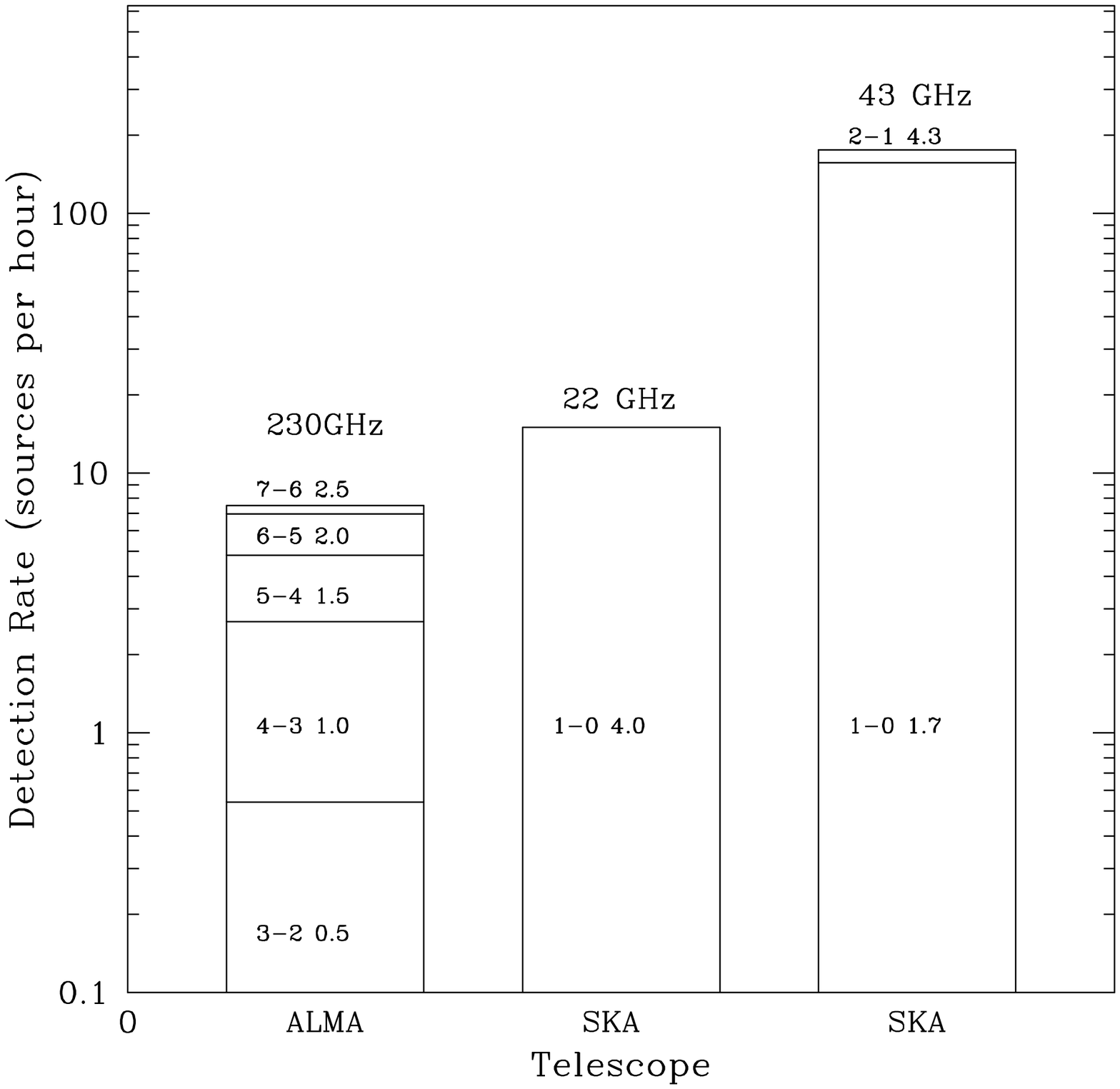,width=6in}
\caption{}
\end{figure}

\clearpage
\newpage

\end{document}